\documentclass{article}
\usepackage{graphicx, amsmath, mathdots} 

\usepackage{geometry}
\usepackage{authblk}
\usepackage{amssymb}
\usepackage{hyperref}

\DeclareMathOperator*{\argmin}{arg\,min}

\def \a {\mathbf{a}}
\def \b {\mathbf{b}}
\def \d {\mathbf{d}}

\def \i {\mathbf{i}}
\def \v {\mathbf{v}}

\def \e {\mathbf{e}}
\def \f {\mathbf{f}}
\def \g {\mathbf{g}}

\def \J {\mathbf{J}}
\def \y {\mathbf{y}}
\def \r {\mathbf{r}}
\def \h {\mathbf{h}}
\def \p {\mathbf{p}}
\def \z {\mathbf{z}}

\def \u {\mathbf{u}}
\def \r {\mathbf{r}}
\def \x {\mathbf{x}}
\def \w {\mathbf{w}}
\def \W {\mathbf{W}}
\def \hx {\hat{\mathbf{x}}}
\def \hf {\hat{\mathbf{f}}}
\def \hg {\hat{\mathbf{g}}}

\def \A {\mathbf{A}}

\def \E {\mathbf{E}}
\def \F {\mathbf{F}}
\def \G {\mathbf{G}}
\def \H {\mathbf{H}}
\def \I {\mathbf{I}}
\def \R {\mathbf{R}}

\def \rT {\mathrm{T}}
\def \rH {\mathrm{H}}
\def \rv {\mathrm{v}}

\def \shsum {\sum_{n=0}^N\sum_{m=-n}^{n}}
\newcommand{\sint}[1]{\int_\mathbf{u} #1 \mathrm{d}\mathbf{u}}
\newcommand{\sintt}[1]{\int_{\mathbf{u}'} #1 \mathrm{d}\mathbf{u}'}

\title{Gaunt coefficients for complex and real spherical harmonics with applications to spherical array processing and Ambisonics}

\author[]{Archontis Politis}

\affil[]{
Technical note v. 1 \\ 
Audio Research Group \\
Faculty of Information Technologies and Communication Sciences \\
Tampere University\\
Finland}

\date{\today}

\begin{document}

\maketitle

\begin{abstract}
Acoustical signal processing of directional representations of sound fields, including source, receiver, and scatterer transfer functions, are often expressed and modeled in the spherical harmonic domain (SHD). Certain such modeling operations, or applications of those models, involve multiplications of those directional quantities, which can also be expressed conveniently in the SHD through coupling coefficients known as Gaunt coefficients. Since the definition and notation of Gaunt coefficients varies across acoustical publications, this work defines them based on established conventions of complex and real spherical harmonics (SHs) along with a convenient matrix form for spherical multiplication of directionally band-limited spherical functions. Additionally, the report provides a derivation of the Gaunt coefficients for real SHs, which has been missing from the literature and can be used directly in spatial audio frameworks such as Ambisonics. Matlab code is provided that can compute all coefficients up to user specified SH orders. Finally, a number of relevant acoustical processing examples from the literature are presented, following the matrix formalism of coefficients introduced in the report.
\end{abstract} \hspace{10pt}

\section{Introduction}

Acoustical signal processing often relies in a directional representation of sound fields, e.g. as a sum of plane waves impinging from various directions observed over some area of interest. Similarly, many acoustical systems have a natural representation as directional functions, e.g. microphone or loudspeaker directivity patterns, head-related transfer functions (HRTFs), or beamforming responses. Acoustical processing in spherical coordinates is a natural choice for such representations, whether it is for spherical array processing \cite{rafaely2015fundamentals, jarrett2017theory}, source-room-receiver interaction \cite{morgenstern2017design}, speech \cite{porschmann2021investigating} or musical instrument directivity modeling \cite{ackermann2021comparative}, head-and-ear-related acoustics \cite{romigh2015efficient}, spatial audio recording \cite{jin2013design, zotter2019ambisonics} and spatial audio rendering \cite{zotter2012all, zotter2019ambisonics}, directional analysis of room acoustics \cite{tervo2015direction}, and spatial audio encoding of room acoustics simulations \cite{bilbao2019local}. 

A common component in all the above examples is that they operate on a spherical harmonic (SH) transform representation of the sound field. It provides a structured way to work with a finite vector representation of the sound field, assess finite approximation errors \cite{kennedy2007intrinsic}, and simplify many operations, such as beamforming \cite{rafaely2015fundamentals} or spatial audio decoding \cite{zotter2019ambisonics}, typically to dot or matrix products of sound field coefficients. Such operations have been covered extensively in literature and in a number of textbooks \cite{rafaely2015fundamentals, jarrett2017theory, zotter2019ambisonics}.

A related, but less common, type of directional processing involves not just spatial filtering, encoding, or decoding, but multiplication of directivity functions or directional representations of sound fields. Such products can be conveniently computed from the SH coefficients of the underlying functions through coupling coefficients known as Gaunt coefficients \cite{sebilleau1998computation}, that are also closely related to Clebsch-Gordan coefficients \cite{ozay2023evaluation}. They appear in a number of disparate studies in acoustical processing works under one name or another, and under varying definitions of complex SHs or real SHs, normalizations, and ordering schemes. In studies working with complex SHs, they have been used e.g. by Shabtai et al. in beamforming fos spherical microphone arrays (SMAs) that preserves binaural cues in \cite{shabtai2013generalized}. 
Kleijn et al. \cite{kleijn2018directional} uses them to derive directional masks that can sharpen peaks and dips of low-order Ambisonics. Since acoustic intensity is a product of the acoustic scalar pressure and vector velocity field, their use appear in a number of works that operate with intensity in the SHD. Zuo et al. \cite{zuo2019spatial} expresses acoustic intensity vector fields directly in terms of SH coefficients of the underlying plane wave amplitude distributions, while Herzog and Habets \cite{herzog2021generalized} formulate spatially-averaged intensity vector estimates for improved directional analysis. Politis and Pulkki \cite{politis2016acoustic} use Gaunt coefficients to perform energy-density and acoustic intensity analysis in angular sectors of the sound field, and subsequently use those parameters for parametric spatial audio rendering of Higher-Order Ambisonics (HOA) in \cite{politis2015sector}. Hold et al. \cite{hold2023compression} uses the same parameterization for spatial audio coding and compression of HOA recordings. They are additionally used by Politis \cite{politis2016diffuse} to derive the diffuse-field coherence between microphones or recording devices of arbitrary directionality - a useful quantity in ambient noise suppression or speech enhancement \cite{bitzer2001superdirective, mccowan2003microphone}. Finally, similar formulations are used to model and estimate parameters of a spatial room impulse response by Bastine et al. \cite{bastine2022time}. 

The most popular representation and format for spatial audio scene recordings or mixes is the Ambisonics format \cite{gerzon1973periphony, zotter2019ambisonics}, which is using the real SH basis functions. Real SH processing of sound scenes has the benefit of producing signals of SH coefficients that are also real, and hence can be stored, processed, transmitted, and listened to as any other normal audio signal. Furthermore, it has a more intuitive connection to recording engineering practice, since it can be seen as a multichannel recording of the sound scene with directivity patterns equivalent to the real SH functions; a popular interpretation for example of first-order Ambisonic recordings. Considering the popularity of Ambisonics and practicality of real SH representations, there are very few works exploiting directly Gaunt coefficients for real SHs. In his thesis Zotter \cite{zotter2009analysis} lists directional windowing as one fundamental operation on Ambisonic recordings. Translation operations of the listening point inside an Ambisonic recording, as investigated e.g. for 6 degrees-of-freedom (6DoF) rendering applications by Tylka et al. \cite{tylka2019domains}, can be also expressed in terms of Gaunt coefficients. Kronnlachner and Zotter \cite{kronlachner2014spatial} use them to apply arbitrary directional loudness modifications through real multiplicative directional masks (even though the coefficients are not used directly but approximated through successive forward-backward discrete SH transforms of the directional products). Lecomte et al. \cite{lecomte2018directional} showcases similar spatial effects. Finally, Alary et al. \cite{alary2019directional} uses them to modify the directional power response of a feedback delay network operating on Ambisonic signals resulting in non-uniform reverberation time across different directions.

Based on the importance of Gaunt coefficients in the above models and applications, the purpose of this report is threefold. a) To present the formulation of the coefficients under a fixed convention of complex SHs and real SHs, serving both the community of sound field modeling, spherical beamforming, and acoustic analysis with SMAs (where complex SHs seem to be more prevalent), and the community of Ambisonics, spatial audio capture and reproduction (where real SH signals are used). Since presentation in acoustical literature is fairly scattered and conventions are not always presented in detail, we hope that the presentation can be useful to the researcher or engineer that tries to study those works. b) Based on the practicality and popularity of Ambisonics and the associated applications, to present a derivation of native Gaunt coefficients for real SHs, following HOA conventions, that can be used directly with Ambisonic recordings for the applications mentioned earlier avoiding transformations to complex SH signals. c) To serve as a reference to Matlab routines that the author has published that can be used to generate matrices of Gaunt coefficients for real or complex SHs. Finally, the report highlights briefly the use of Gaunt coefficients in some of the aforementioned published applications, using matrices of Gaunt coefficients for real SHs, as derived in this report.

\section{Preliminaries}

The complex spherical harmonics of order $n$ and degree $m$ are defined following the convention
\begin{align}
    Y_{n,m}(\theta,\phi) &= (-1)^m\sqrt{\frac{(2n+1)(n-m)!}{4\pi(n+m)!}}P_{n,m}(\cos\theta)e^{im\phi} \nonumber\\
    &= (-1)^m \Theta_{n,m}(\theta)e^{im\phi}
      \label{eq:complexSH}  
\end{align}
with integers $n\geq0$ and $-n\leq m \leq n$ being termed \emph{order} and \emph{degree} respectively, and including the Condon-Shortley phase term $(-1)^m$. The (unnormalized) associated Legendre functions $P_{n,m}$ are defined as
(omitting the Condon-Shortley phase term from their definition, included instead in the definition of SHs)
\begin{equation}
    P_{n,m}(x) = (1-x^2)^{m/2} \frac{d^m}{dx^m} P_n(x)
\end{equation}
for $0\leq m\leq n$, with $P_n$ being the Legendre polynomials
\begin{equation}
    P_n(x) = \frac{1}{2^n n!}\frac{d^n}{d x^n}(x^2-1)^n.
\end{equation}
The complex SHs of negative orders $-n \leq m<0 $ can then be computed through the property 
\begin{equation}
    Y_{n,m}^*(\theta,\phi) = (-1)^m Y_{n,-m}(\theta,\phi). \label{eq:conjSH}
\end{equation}
Real spherical harmonics can be defined with respect to the complex ones as
\begin{align}
    R_{n,m}(\theta,\phi) = \begin{cases}
        \frac{(-1)^m}{\sqrt{2}}(Y_{n,m}(\theta,\phi) + Y_{n,m}^*(\theta,\phi)), & \mathrm{for}\; m>0\\
Y_{n,m}(\theta,\phi), & \mathrm{for}\; m=0\\
        \frac{(-1)^m}{i\sqrt{2}}(Y_{n,|m|}(\theta,\phi) - Y_{n,|m|}^*(\theta,\phi)), & \mathrm{for}\; m<0\\
    \end{cases},
    \label{eq:realSH1}
\end{align}
where $()^*$ denotes conjugation. Based on Eq.~ \ref{eq:conjSH}, Eq.~\ref{eq:realSH1} becomes also
\begin{align}
    R_{n,m}(\theta,\phi) = \begin{cases}
        \frac{1}{\sqrt{2}}((-1)^m Y_{n,m}(\theta,\phi) + Y_{n,-m}(\theta,\phi)), & \mathrm{for}\; m>0\\
Y_{n,m}(\theta,\phi), & \mathrm{for}\; m=0\\
        \frac{1}{i\sqrt{2}}((-1)^m Y_{n,-m}(\theta,\phi) - Y_{n,m}(\theta,\phi)), & \mathrm{for}\; m<0\\
    \end{cases}.
        \label{eq:realSH2}
\end{align}
Finally, again based on Eq.~ \ref{eq:conjSH} and Eq.~\ref{eq:realSH1}, they can be defined only in terms of conjugated complex SHs, a form that will prove useful later
\begin{align}
    R_{n,m}(\theta,\phi) = \begin{cases}
        \frac{1}{\sqrt{2}}( Y_{n,-m}^*(\theta,\phi) + (-1)^m Y_{n,m}^*(\theta,\phi)), & \mathrm{for}\; m>0\\
Y_{n,m}(\theta,\phi)), & \mathrm{for}\; m=0\\
        \frac{1}{i\sqrt{2}}(Y_{n,m}^*(\theta,\phi) - (-1)^m Y_{n,-m}^*(\theta,\phi)), & \mathrm{for}\; m<0\\
    \end{cases}.
    \label{eq:realSH3}
\end{align}
Note that in this definition of real SHs, the Condon-Shortley phase term is canceled out during conversion from complex SHs. This form is common in various domains using real SHs, see e.g. \cite{sloan2023precomputed, blanco1997evaluation}, including spatial audio processing following the formalism of \emph{Ambisonics} \cite{gerzon1973periphony, daniel2000acoustic, zotter2019ambisonics}. 

A direct form of the real SHs is then
\begin{align}
    R_{n,m}  &= \sqrt{\frac{(2n+1)(n-|m|)!}{4\pi(n+|m|)!}}P_{n,|m|}(\cos\theta)\Phi_m(\phi) \nonumber\\
            &= \Theta_{n,|m|}(\theta)\Phi_m(\phi)
\end{align}
with 
\begin{equation}
    \Phi_m(\phi) =
\begin{cases}
\sqrt{2}\cos(m \phi), & \mathrm{for}\; m>0\\
1, &  \mathrm{for }\; m=0\\
\sqrt{2}\sin(|m|\phi), & \mathrm{for }\; m<0
\end{cases}.
\label{eq:realSH}
\end{equation}

We can stack the set of $(2n+1)$ SHs of a certain order $n$ in a vector 
\begin{equation}
\bar{\mathbf{y}}_{n}(\theta,\phi) = \left[
\begin{array}{c}
     Y_{n,-n}(\theta,\phi)\\
     .\\
    .\\
    .\\
    Y_{n,0}(\theta,\phi)\\
     .\\
    .\\
    .\\
    Y_{n,n}(\theta,\phi)
\end{array}
\right], \;\mathrm{and}\; 
\bar{\mathbf{r}}_{n}(\theta,\phi) = \left[
\begin{array}{c}
     R_{n,-n}(\theta,\phi)\\
     .\\
    .\\
    .\\
    R_{n,0}(\theta,\phi)\\
     .\\
    .\\
    .\\
    R_{n,n}(\theta,\phi)
\end{array}
\right]
\end{equation}
The two sets of basis functions are connected by unitary matrices
\begin{align}
    \bar{\mathbf{r}}_{n}(\theta,\phi) &= \bar{\mathbf{U}}_n^{(\mathrm{c2r})} \bar{\mathbf{y}}_{n}^{}(\theta,\phi) \label{eq:Uc2r}\\
        \bar{\mathbf{y}}_{n}(\theta,\phi) &= \bar{\mathbf{U}}_n^{(\mathrm{r2c})} \bar{\mathbf{r}}_{n}^{}(\theta,\phi)
        \label{eq:Ur2c}
\end{align}
with 
\begin{align}
\bar{\mathbf{U}}_n^{(\mathrm{c2r})} &= (\bar{\mathbf{U}}_n^{(\mathrm{r2c})})^{-1} = (\bar{\mathbf{U}}_n^{(\mathrm{r2c})})^\mathrm{H}  \\
     \bar{\mathbf{U}}_n^{(\mathrm{r2c})} &= (\bar{\mathbf{U}}_n^{(\mathrm{c2r})})^{-1} = (\bar{\mathbf{U}}_n^{(\mathrm{c2r})})^\mathrm{H}. 
\end{align}
To simplify the notation we use from now on a single $\bar{\mathbf{U}}_n^{} = \bar{\mathbf{U}}_n^{(\mathrm{c2r})}$, with $\bar{\mathbf{U}}_n^\mathrm{H}$ replacing $\bar{\mathbf{U}}_n^{(\mathrm{r2c})}$ if needed.
The structure of the $\bar{\mathbf{U}}_n^{}$ matrices, according to the above conventions and Eq.~\ref{eq:realSH1}-\ref{eq:realSH3}, is
\begin{equation*}
\left[
\begin{array}{l}
     R_{n,-n}(\theta,\phi)\\
     \vdots\\
     R_{n,-|m|}(\theta,\phi)\\
     \vdots\\     
    R_{n,-1}(\theta,\phi)\\
    R_{n,0}(\theta,\phi)\\
     R_{n,1}(\theta,\phi)\\
    \vdots\\
     R_{n,m}(\theta,\phi)\\
     \vdots\\    
    R_{n,n}(\theta,\phi)
\end{array}
\right] =  \frac{1}{\sqrt{2}}
\left[
\begin{array}{ccccccccccc}
i       &           	&           &           	&    	&          	&      &           	&       	&          	& -i(-1)^n \\
        & \ddots    &           &           	&    	&          	&      &           	&       	& \iddots  	&    \\
        &           	& i         &           	&    	&          	&      &           	& -i(-1)^m &          	&    \\
        &           	&           & \ddots    	&    	&          	&      & \iddots   &       	&          	&    \\       
        &           	&           &           	&  i 	&          	&  i   &           	&       	&          	&    \\   
        &           	&           &           	&    	& \sqrt{2} 	&      &           	&       	&          	&    \\        
        &           	&           &           	&  1 	&          	&  1	&           	&       	&          	&    \\ 
        &           	&           & \iddots   	&    	&          	&      & \ddots    &       	&          	&    \\        
        &           	& 1        &           	&    	&          	&      &           	& (-1)^m 	&          	&    \\
        & \iddots   &           &           	&    	&          	&      &           	&       	& \ddots   	&    \\
1      &           	&           &           	&    	&          	&      &           	&       	&          	& (-1)^n    
\end{array}
\right] \cdot
\left[
\begin{array}{l}
     Y_{n,-n}(\theta,\phi)\\
     \vdots\\
     Y_{n,-|m|}(\theta,\phi)\\
     \vdots\\     
    Y_{n,-1}(\theta,\phi)\\
    Y_{n,0}(\theta,\phi)\\
     Y_{n,1}(\theta,\phi)\\
    \vdots\\
     Y_{n,m}(\theta,\phi)\\
     \vdots\\    
    Y_{n,n}(\theta,\phi)
\end{array}
\right].
\end{equation*}

According to the property $Y_{n,m}^*(\theta,\phi) = (-1)^m Y_{n,-m}(\theta,\phi)$, we can similarly define a transformation matrix from a vector of complex SHs to a vector of complex conjugated SHs
\begin{equation}
    \bar{\mathbf{y}}_{n}^*(\theta,\phi) = \bar{\mathbf{T}}_n \bar{\y}_n(\theta,\phi)
    \label{eq:Tn}
\end{equation}
where $\bar{\mathbf{y}}_{n}^*(\theta,\phi) = [Y_{n,-n}^*(\theta,\phi),...,Y_{n,n}^*(\theta,\phi)]^\rT$ and $\bar{\mathbf{T}}_n$ is a symmetric anti-diagonal matrix with \\$[(-1)^n,(-1)^{n-1},...,1,...,(-1)^{n-1},(-1)^n]^\rT$ in the anti-diagonal.

It is convenient when dealing with directionally band-limited functions of some maximum order $N$ to handle vectors of all SHs up to that order 
$\mathbf{y}_N(\theta,\phi) = [\bar{\mathbf{y}}_0^\mathrm{T}(\theta,\phi),...,\bar{\mathbf{y}}_n^\mathrm{T}(\theta,\phi),...,\bar{\mathbf{y}}_N^\mathrm{T}(\theta,\phi)]^\mathrm{T}$. $\mathbf{r}_N(\theta,\phi)$ is similarly defined. The transformation matrices of these vectors from complex to real are easily constructed from the order-specific ones of Eq.~\ref{eq:Uc2r}-\ref{eq:Ur2c} as block diagonal matrices
\begin{align}
    \mathbf{U}_N = \left[\begin{array}{ccccc}
      \bar{\mathbf{U}}_0   &       &                                   &       & \\
                                        & \ddots &                                   &      & \\
                                        &       & \bar{\mathbf{U}}_n   &       &\\
                                        &       &                                   & \ddots &\\
                                        &       &                                   &       & \bar{\mathbf{U}}_N
    \end{array}\right].
\end{align}
Based on these definitions
\begin{align}
    \mathbf{r}_N(\theta,\phi) &= \mathbf{U}_N \,  \mathbf{y}_N(\theta,\phi)
    \label{eq:c2r}\\
    \mathbf{y}_N(\theta,\phi) &= \mathbf{U}_N^\mathrm{H} \,
\mathbf{r}_N^{}(\theta,\phi)
\label{eq:r2c}
\end{align}
and, based on Eq.~\ref{eq:Tn}, we can similarly define
\begin{align}
    \mathbf{y}_{N}^*(\theta,\phi) = \mathbf{T}_N^{} \y_N^{}(\theta,\phi)
    \label{eq:TN}
\end{align}
with $\mathbf{T}_N$ constructed from $\bar{\mathbf{T}}_n$ similarly to $\mathbf{U}_N$.

\section{Spherical Harmonic Transform}

We define the spherical harmonic transform (SHT) of a spherical function $f(\mathbf{u})$, directionally band-limited up to order $N$, as
\begin{align}
      \mathbf{f}_{N}^\rT & = c\mathcal{SHT}_N[f(\mathbf{u})] = \int_{\mathbf{u}\in\mathcal{S}^2} f(\mathbf{u})\, \y_{N}^\rH(\mathbf{u}) \mathrm{d}\mathbf{u} \\
      \hat{\mathbf{f}}_{N}^\rT & =  r\mathcal{SHT}_N[f(\mathbf{u})] = \int_{\mathbf{u}\in\mathcal{S}^2} f(\mathbf{u})\, \r_{N}^\rT(\mathbf{u}) \mathrm{d}\mathbf{u}    
\end{align}
where, for compactness, we define the direction unit vector $\mathbf{u}$ in the direction of $(\theta,\phi)$ and spherical integration as $\int_{\mathbf{u}\in\mathcal{S}^2} (\cdot) \mathrm{d}\mathbf{u} = \int_{\phi=0}^{2\pi}\int_{\theta=0}^{\pi/2} (\cdot) \sin\theta\mathrm{d}\theta \mathrm{d}\phi$. $\mathbf{f}_N$ and $\hat{\mathbf{f}}_N$ are the SHT coefficients of $f(\mathbf{u})$ projected on complex and real SHs respectively. The vectors of SHT coefficients are following the same indexing as the SHs, e.g. $\mathbf{f}_N = [f_{0,0}, f_{-1,1},...,f_{n,m},...,f_{N,N}]^\mathrm{T}$.

The inverse SHT is then given by
\begin{align}
      f(\mathbf{u}) &= \sum_{n=0}^N\sum_{m=-n}^n f_{n,m}Y_{n,m}(\mathbf{u}) = \mathbf{f}_{N}^\mathrm{T}\, \y_{N}^{}(\mathbf{u}) \label{eq:ishtc} \\
       f(\mathbf{u}) &= \sum_{n=0}^N\sum_{m=-n}^n \hat{f}_{n,m}R_{n,m}(\mathbf{u}) = \hat{\mathbf{f}}_{N}^\mathrm{T}\, \r_{N}^{}(\mathbf{u}).
      \label{eq:ishtr}
\end{align}
From Eq.~\ref{eq:ishtc}-\ref{eq:ishtr} and Eq.~\ref{eq:c2r}-\ref{eq:r2c} we can get directly the conversion of SHT coefficients between the complex and real basis
\begin{align}
      \hat{\mathbf{f}}_{N}^\rT &= \mathbf{f}_{N}^\rT \, \mathbf{U}_N^\rH \\
      \mathbf{f}_{N}^\rT &= \hat{\mathbf{f}}_{N}^\rT \, \mathbf{U}_N^{}.
\end{align}
Another property of interest is the relation between the SH coefficients of a conjugated function $f^*(\u)$ with respect to the coefficients $\f_N$ or $\hf_N$ of the original $f(\u)$
\begin{align}
      r\mathcal{SHT}_N[f^*(\mathbf{u})] &= \hf_N^\rH \\
      c\mathcal{SHT}_N[f^*(\mathbf{u})] &= \f_N^\rH \mathbf{T}_N^{},
      \label{eq:conjfcoeffs}
\end{align}
where the second one is derived using Eq.~\ref{eq:TN}.

Due to the orthonormality of the complex and real SHs, it is also straightforward to demonstrate that for two functions $f(\u)$ and $g(\u)$ band limited to order $N$, with coefficients $\f_N, \g_N$ or $\hf_N, \hg_N$ the following holds
\begin{align}
      \sint{ |f(\u)|^2 } &= ||\f_N||_2^2 =  ||\hf_N||_2^2 \\
      \sint{ f^*(\u) g(\u) } &= \f_N^\rH \g_N =  \hf_N^\rH \hg_N^{}.
\end{align}
Additionally, 
\begin{align}
      \sint{ f(\u) g(\u) } &= \hf_N^\rT \hg_N^{},
\end{align}
in terms of real SHs, while, in terms of complex SHs and based on Eq.~\ref{eq:conjfcoeffs}
\begin{align}
      \sint{ f(\u) g(\u) } =  \sint{ (f^*(\u))^* g(\u) } = \f_N^\rT \mathbf{T}_N^{} \g_N^{}.
\end{align}

\section{Spherical multiplication}

The central operation of interest that leads to computation of Gaunt coefficients is multiplication of two spherical functions $h(\mathbf{u})=f(\mathbf{u})\cdot g(\mathbf{u})$, encountered in various applications. Since we may have a model of the factors $f(\mathbf{u})$ and $g(\mathbf{u})$ in terms of their SHT coefficients, it may be convenient to have a description of the product also in terms of SHT coefficients. In that case, and assuming that the factors are band-limited to some order $N'$ and $N''$, we are looking to express the coefficients of the product $\mathbf{h}_N$ in terms of the coefficients of the factors $\mathbf{f}_{N'}, \mathbf{g}_{N''}$. Since the factors are band-limited the product will be also band-limited to order $N=N'+N''$.

Focusing at a single coefficient of the product function using the complex SH basis, we obtain
\begin{align}
  h_{n,m} &=    \int_\mathbf{u} h(\mathbf{u}) Y_{n,m}^*(\mathbf{u}) \mathrm{d}\mathbf{u} 
  =    \int_\mathbf{u} f(\mathbf{u})g(\mathbf{u}) Y_{n,m}^*(\mathbf{u}) \mathrm{d}\mathbf{u} \nonumber\\
  &=    \int_\mathbf{u} (\mathbf{f}_{N'}^\mathrm{T} \y_{N'}^{}(\mathbf{u})) (\mathbf{g}_{N''}^\mathrm{T} \y_{N''}^{}(\mathbf{u})) Y_{n,m}^*(\mathbf{u}) \mathrm{d}\mathbf{u} \nonumber\\
  &=   \mathbf{f}_{N'}^\mathrm{T} \left( \int_\mathbf{u}  \y_{N'}(\mathbf{u}) \y_{N''}^\mathrm{T}(\mathbf{u}) Y_{n,m}^*(\mathbf{u}) \mathrm{d}\mathbf{u}\right)\, \mathbf{g}_{N''}^{} \nonumber\\   
  &=   \mathbf{f}_{N'}^\mathrm{T} \mathbf{G}_{N',N''}^{n,m} \mathbf{g}_{N''}^{}.
  \label{eq:cGmtx}
\end{align}
Hence, the SHT of the product reduces to a bilinear form between the coefficients of the two factors and certain matrices of integrals of SHs that couple the two sets of coefficients. Those matrices $\mathbf{G}_{N',N''}^{n,m}$ of size $(N'+1)^2)\times(N''+1)^2$ contain Gaunt coefficients $G_{n',m',n'',m''}^{n,m} $ with the following indexing $[\mathbf{G}_{N',N''}^{n,m}]_{q,l}$
\begin{align}
    n' &= \lfloor\sqrt{q-1}\rfloor \quad\mathrm{and}\quad m' = q-(n')^2-(n')-1 \nonumber\\
    n'' &= \lfloor\sqrt{l-1}\rfloor \quad\mathrm{and}\quad m'' = l-(n'')^2-(n'')-1,
\end{align}
with $\lfloor\cdot\rfloor$ indicating the floor function.
They can be precomputed up to dimensions determined by the maximum practical orders $N',N''$ involved in a certain application of interest, with $n\leq N'+N''$.

\section{Gaunt coefficients for complex SHs}

The Gaunt coefficient $G_{n',m',n'',m''}^{n,m}$ for complex SHs is defined as \cite{sebilleau1998computation}
\begin{align}
    G_{n',m',n'',m''}^{n,m} = \int_{\mathbf{u}\in\mathcal{S}^2} Y_{n',m'}(\mathbf{u}) Y_{n'',m''}(\mathbf{u}) Y_{n,m}^*(\mathbf{u}) \mathrm{d}\mathbf{u}
\end{align}
Essentially, the coefficient expresses the SHT of the product of two complex SHs
\begin{align}
     Y_{n',m'}(\mathbf{u}) Y_{n'',m''}(\mathbf{u}) = \sum_n \sum_m G_{n',m',n'',m''}^{n,m} Y_{n,m}(\mathbf{u}).
\end{align}
They are real and they are equal to zero unless two conditions are met: $|n'-n''|<n<n'+n''$ and $m=m'+m''$. They also satisfy various symmetry relations which can be taken into account to speed up computations, see e.g. \cite{sebilleau1998computation}. They are directly related to Clebsh-Gordan coefficients \cite{ozay2023evaluation} and can be computed with a number of fast algorithms \cite{clercx1993alternative, xu1996fast, sebilleau1998computation}. A direct formula for their calculation is introduced by Cruzan \cite{cruzan1962translational}
\begin{align}
   G_{n',m',n'',m''}^{n,m} &= (-1)^m\sqrt{\frac{(2n+1)(2n'+1)(2n''+1)}{4\pi}} \left(\begin{array}{ccc}
         n& n' & n''   \\
         0& 0 & 0
    \end{array} \right)
    \left( \begin{array}{ccc}
         n& n' & n''   \\
         -m& m' & m''
    \end{array} \right) \label{eq:Cruzan}
\end{align}
with $\left( \begin{array}{ccc}
         n& n' & n''   \\
         m& m' & m''
    \end{array} \right)$ being Wigner 3-j symbols, which can in turn be computed by e.g. Racah's formula \cite[\href{https://dlmf.nist.gov/34.2}{(34.2.4)}]{NIST:DLMF}.

\section{Gaunt coefficients for real SHs}

We define the Gaunt coefficient $F_{n',m',n'',m''}^{n,m}$ for real SHs similarly to the Gaunt coefficient for complex SHs
\begin{align}
    F_{n',m',n'',m''}^{n,m} = \int_{\mathbf{u}\in\mathcal{S}^2} R_{n',m'}(\mathbf{u}) R_{n'',m''}(\mathbf{u}) R_{n,m}(\mathbf{u}) \mathrm{d}\mathbf{u}
\end{align}
expressing the SHT of the product of two real SHs
\begin{align}
     R_{n',m'}(\mathbf{u}) R_{n'',m''}(\mathbf{u}) = \sum_n \sum_m F_{n',m',n'',m''}^{n,m} R_{n,m}^{}(\mathbf{u}).\label{eq:2shprod}
\end{align}
Focusing at a single coefficient of the product function using real SH basis, and following the same steps as Eq.~\ref{eq:cGmtx}
\begin{align}
  \hat{h}_{n,m} &=    \int_\mathbf{u} h(\mathbf{u}) R_{n,m}(\mathbf{u}) \mathrm{d}\mathbf{u} \nonumber\\
    &= \hat{\mathbf{f}}_{N'}^\mathrm{T} \left( \int_\mathbf{u}  \r_{N'}^{}(\mathbf{u}) \r_{N''}^\mathrm{T}(\mathbf{u}) R_{n,m}(\mathbf{u}) \mathrm{d}\mathbf{u}\right)\, \hat{\mathbf{g}}_{N''}^{} \nonumber\\
  &= \hat{\mathbf{f}}_{N'}^\mathrm{T} \mathbf{F}_{N',N''}^{n,m} \hat{\mathbf{g}}_{N''}^{}.
  \label{eq:rGmtx}
\end{align}
The matrices  $\mathbf{F}_{N',N''}^{n,m}$ contain Gaunt coefficients for real SHs indexed in the same manner as in the complex SH case. These coefficients and their properties are studied by e.g. Homeier and Steinborn \cite{homeier1996some}. They can be straightforwardly obtained in terms of their complex SH counterparts and the transformation matrices $\mathbf{U}_N$ from complex to real SHs 
\begin{align}
 \mathbf{F}_{N',N''}^{n,m} &= \int_\mathbf{u}  \r_{N'}^{}(\mathbf{u}) \r_{N''}^\mathrm{T}(\mathbf{u}) R_{n,m}(\mathbf{u}) \mathrm{d}\mathbf{u}  \nonumber\\  
 &= \int_\mathbf{u}  \mathbf{U}_{N'}^{}\y_{N'}^{}(\mathbf{u}) \y_{N''}^\mathrm{T}(\mathbf{u})\mathbf{U}_{N''}^\mathrm{T} R_{n,m}(\mathbf{u}) \mathrm{d}\mathbf{u}  \nonumber\\  
  &= \mathbf{U}_{N'}^{} \left(\int_\mathbf{u}  \y_{N'}^{}(\mathbf{u}) \y_{N''}^\mathrm{T}(\mathbf{u}) R_{n,m}(\mathbf{u}) \mathrm{d}\mathbf{u} \right) \mathbf{U}_{N''}^\mathrm{T} . 
\end{align}
Using Eq.~\ref{eq:realSH3} we get
\begin{align}
 \mathbf{F}_{N',N''}^{n,m} &= \mathbf{U}_{N'} \left(\int_\mathbf{u}  \y_{N'}^{}(\mathbf{u}) \y_{N''}^\mathrm{T}(\mathbf{u}) R_{n,m}(\mathbf{u}) \mathrm{d}\mathbf{u} \right) \mathbf{U}_{N''}^\mathrm{T}  \nonumber\\  
 &= \begin{cases}
\mathbf{U}_{N'}^{} \left(\int_\mathbf{u}  \y_{N'}^{}(\mathbf{u}) \y_{N''}^\mathrm{T}(\mathbf{u}) \frac{1}{\sqrt{2}}( Y_{n,-m}^*(\theta,\phi) + (-1)^m Y_{n,m}^*(\theta,\phi)) \mathrm{d}\mathbf{u} \right) \mathbf{U}_{N''}^\mathrm{T}& \mathrm{for}\; m>0 \\
\mathbf{U}_{N'}^{} \left(\int_\mathbf{u}  \y_{N'}^{}(\mathbf{u}) \y_{N''}^\mathrm{T}(\mathbf{u}) Y_{n,0}^*(\mathbf{u}) \mathrm{d}\mathbf{u} \right) \mathbf{U}_{N''}^\mathrm{T}& \mathrm{for}\; m=0 \\
\mathbf{U}_{N'}^{} \left(\int_\mathbf{u}  \y_{N'}^{}(\mathbf{u}) \y_{N''}^\mathrm{T}(\mathbf{u}) \frac{1}{i\sqrt{2}}(Y_{n,m}^*(\theta,\phi) - (-1)^m Y_{n,-m}^*(\theta,\phi)) \mathrm{d}\mathbf{u} \right) \mathbf{U}_{N''}^\mathrm{T}& \mathrm{for}\; m<0
 \end{cases}. 
\end{align}
Based on the final result, it is straightforward to obtain the Gaunt coefficient matrices for real SHs with respect to the ones for complex SHs:
\begin{align}
 \mathbf{F}_{N',N''}^{n,m} = \begin{cases}
\frac{1}{\sqrt{2}} \mathbf{U}_{N'}^{} \left(\mathbf{G}_{N',N''}^{n,-m} + (-1)^m  \mathbf{G}_{N',N''}^{n,m} \right) \mathbf{U}_{N''}^\mathrm{T}& \mathrm{for}\; m>0 \\
\mathbf{U}_{N'}^{}  \mathbf{G}_{N',N''}^{n,0} \mathbf{U}_{N''}^\mathrm{T}& \mathrm{for}\; m=0 \\
\frac{1}{i\sqrt{2}} \mathbf{U}_{N'}^{} \left( \mathbf{G}_{N',N''}^{n,m} - (-1)^m \mathbf{G}_{N',N''}^{n,-m}) \right) \mathbf{U}_{N''}^\mathrm{T}& \mathrm{for}\; m<0
 \end{cases}. \label{eq:cGmtx2rGmtx}
\end{align}

\section{Applications to spherical array processing and Ambisonics}

We show some examples of applications of matrices of Gaunt coefficients in spherical acoustics, spatial audio, and Ambisonics. Similar derivations have appeared in the literature but without a common convention, and often through related quantities (e.g. Clebsh-Gordan coefficients). Moreover, typically Gaunt coefficients have been used in acoustical and audio research with complex SHs, leaving out popular conventions such as Ambisonics. In the following examples we follow use of Gaunt coefficients for real SHs, even though most of the derivations remain similar when switching to complex SHs.

\subsection{Sound field translation}

Consider a sound field characterized by a plane wave density $a(k,\u)$ that is directionally band-limited to order $N$, so that it is characterized by the vector of SH coefficients $\a_N(k) = r\mathcal{SHT}_N[a(k,\u)]$. The acoustic pressure at a position $\x = (\theta',\phi',d)$ around the origin is given by the finite Fourier-Bessel expansion
\begin{equation}
    p(k,\x) = \shsum 4\pi i^n j_n(kd)a_{n,m}(k) R_{n,m}(\hx) = \a_N^\rT(k) \J_N^{}(kd) \r_N^{}(\hx)
    \label{eq:fbe}
\end{equation}
where $j_n$ is the spherical Bessel function of order $n$ and $\hx=\x/||\x||$ is the direction vector of $\x$. The diagonal matrix $\J_N$ contains the $4\pi i^n j_n(kd)$ terms indexed as $[\J_N]_{q,q} = 4\pi i^{\lfloor \sqrt{q-1} \rfloor} j_{\lfloor \sqrt{q-1} \rfloor}(kd)$ with $\lfloor \cdot \rfloor$ being the floor function. Similarly, the case that the sound field is composed only from a single unit-amplitude plane wave incident from $\u$ results in
\begin{align}
    e^{ik\u^\rT\x} &\approx \sum_{n''=0}^{N''} \sum_{m''=-n''}^{n''} 4\pi i^{n''} j_{n''}(kd)  R_{n'',m''}(\u) R_{n'',m''}(\hx) \nonumber\\
    & = \r_{N''}^\rT(\u) \J_{N''}^{}(kd) \r_{N''}^{}(\hx),
    \label{eq:pwe}
\end{align}
where the plane wave expansion can be truncated to a suitable order $N''$ with negligible error for the application at hand, following e.g. a rule of $N''=\lceil ekd/2 \rceil $ \cite{kennedy2007intrinsic}.

Now let us consider a new origin at a translation $\x$ with the respect to the original origin. The translated plane-wave amplitude density at that position is $a_\mathrm{t}(k,\u,\x) = a(k,\u) e^{ik\u^\rT\x}$. For various applications such as sound field interpolation/extrapolation for 6-degrees-of-freedom rendering \cite{tylka2020performance, zotter2009analysis}, or sound field interpolation from distributed spherical microphone arrays (SMAs) \cite{samarasinghe2014wavefield, ueno2017sound} it is convenient to be able to express the sound field SH coefficients $\a_{N'}^{(t)}$ at the translated origin with respect to the coefficients at the original origin $\a_N$. 

Taking the SHT of $a_\mathrm{t}(k,\u,\x)$ and using Eq.~\ref{eq:pwe} we have
\begin{align}
    a_{n',m'}^{(t)}(k,\x) &= \sint{a_\mathrm{t}(k,\u,\x)R_{n',m'}(\u)}
    = \sint{a(k, \u)e^{ik\u^\rT \x} R_{n',m'}(\u)} \nonumber\\
    &= \sint{ \left(\a_N^\rT(k)\r_N^{}(\u)\right) \left( \r_{N''}^\rT(\u) \J_{N''}^{}(kd) \r_{N''}^{}(\hx) \right) R_{n',m'}(\u)  } \nonumber\\
    &= \a_N^\rT(k) \left(\sint{ \r_N^{}(\u) \r_{N''}^\rT(\u) R_{n',m'}(\u) } \right) \J_{N''}^{}(kd) \r_{N''}^{}(\hx) \nonumber\\
    &= \a_N^\rT(k) \G_{N,N''}^{n',m'} \J_{N''}^{}(kd) \r_{N''}^{}(\hx).
\end{align}
Using the above finite order approximations, the translated sound field coefficients $\a_{N'}^{(t)}$ are order limited to $N'=N+N''$.

\subsection{Acoustic intensity fields}

Acoustic intensity is a measurable quantity using microphone \cite{hacihabibouglu2013theoretical}, or other sensor \cite{de2003overview}, arrays with various applications in acoustics \cite{tervo2009direction, merimaa2005spatial, fahy2017sound}. In the context of spatial audio processing, it has a central role as a suitable parameter for sound field coding and parameterization and rendering, as used e.g. in DirAC \cite{pulkki2017first, politis2015sector, herzog2021generalized}, and as a proxy for the perceived localization of a spatial audio reproduction system, as used e.g. in \cite{grandjean2018size, de2013analysis}. For known, or measured sound fields, characterized by some directional amplitude density $a(\u)$, it can be useful to describe the acoustic intensity field at some point around the origin $\x$ in terms of the SH coefficients up to order $N$ with $\a_N = r\mathcal{SHT}_N[a(\u)]$. The narrowband complex acoustic intensity vector $\i(\x,f)$ is given by the product of the acoustic pressure $p(\x,f)$ and velocity field $\v(\x,f)$
\begin{equation}
    \i(\x,k) = \frac{1}{2}p^*(\x,k) \v(\x,k).
    \label{eq:acintvec}
\end{equation}

The acoustic particle velocity vector $\v$ relates to the plane wave density as
\begin{align}
    \v(\x,k) &= \sint{ -\frac{1}{c\rho_0} a(k, \u) e^{ik\u^\rT\x} \u } \nonumber\\
            &= -\frac{1}{c\rho_0} \sint{  (\a_N^\rT(k)\r_N^{}(\u)) ( \r_{N''}^\rT(\u) \J_{N''}^{}(kd) \r_{N''}^{}(\hx) ) \u } 
\end{align}
The cartesian components of $\u$ are directly linked to either a combination of first-order complex SHs, or directly the first order real SHs
\begin{align}
    \u = \left[ \begin{array}{c}
         \sin\theta\cos\phi  \\
         \sin\theta\sin\phi \\
         \cos\theta         
    \end{array} \right] =
    \sqrt{\frac{4\pi}{3}} 
        \left[ \begin{array}{c}
         R_{1,1}(\u)  \\
         R_{1,-1}(\u) \\
         R_{1,0}(\u)
    \end{array} \right] =
    \sqrt{\frac{4\pi}{6}} 
        \left[ \begin{array}{c}
         -Y_{1,1}(\u) + Y_{1,-1}(\u) \\
         -i Y_{1,1}(\u) + i Y_{1,-1}(\u) \\
         \sqrt{2} Y_{1,0}(\u)
    \end{array} \right].    
\end{align}
We focus on the presentation of the velocity vector using the real SH form of $\u$
\begin{align}
    \left[ \begin{array}{c}
            \rv_x(\x,k)\\
           \rv_y(\x,k)\\
           \rv_z(\x,k)
    \end{array}
    \right]  &= -\sqrt{\frac{4\pi}{3}}\frac{1}{c\rho_0}\left[\begin{array}{c}
    \sint{ (\a_N^\rT(k)\r_N^{}(\u)) ( \r_{N''}^\rH(\u) \J_{N''}^{}(kd) \r_{N''}^{}(\hx) ) R_{1,1}(\u) }\\
    \sint{ (\a_N^\rT(k)\r_N^{}(\u)) ( \r_{N''}^\rH(\u) \J_{N''}^{}(kd) \r_{N''}^{}(\hx) ) R_{1,-1}(\u)  }\\
    \sint{ (\a_N^\rT(k)\r_N^{}(\u)) ( \r_{N''}^\rH(\u) \J_{N''}^{}(kd) \r_{N''}^{}(\hx) ) R_{1,0}(\u) }       
    \end{array} 
    \right] \nonumber\\
&= -\sqrt{\frac{4\pi}{3}}\frac{1}{c\rho_0}
    \left[\begin{array}{c}
    \a_N^\rT(k)\left(\sint{ \r_N^{}(\u)) \r_{N''}^\rT(\u)R_{1,1}(\u) } \right)\J_{N''}^{}(kd) \r_{N''}^{}(\hx) \\
    \a_N^\rT(k)\left(\sint{ \r_N^{}(\u)) \r_{N''}^\rT(\u) R_{1,-1}(\u) }\right)\J_{N''}^{}(kd) \r_{N''}^{}(\hx) \\
    \a_N^\rT(k)\left(\sint{ \r_N^{}(\u)) \r_{N''}^\rT(\u) R_{1,0}(\u) }\right) \J_{N''}^{}(kd) \r_{N''}^{}(\hx)    
    \end{array} 
    \right]    \nonumber\\  
&= -\sqrt{\frac{4\pi}{3}}\frac{1}{c\rho_0}
     \left[\begin{array}{c}
    \a_N^\rT(k)\F_{N,{N''}}^{1,1} \J_{N''}^{}(kd) \r_{N''}^{}(\hx) \\
    \a_N^\rT(k)\F_{N,{N''}}^{1,-1} \J_{N''}^{}(kd) \r_{N''}^{}(\hx)\\
    \a_N^\rT(k)\F_{N,{N''}}^{1,0} \J_{N''}^{}(kd) \r_{N''}^{}(\hx)    
    \end{array} 
    \right]       
\end{align}
Using Eq.~\ref{eq:acintvec} \& Eq.~\ref{eq:fbe} the acoustic intensity vector becomes
\begin{align}
    \i(\x,k) &= \frac{1}{2}p^*(\x,k) \v(\x,k) \nonumber\\
            &= \frac{1}{2}\left( \a_N^\rT(k) \J_N^{}(kd) \r_N^{}(\hx) \right)^\rH \left( -\sqrt{\frac{4\pi}{3}}\frac{1}{c\rho_0}
     \left[\begin{array}{c}
    \a_N^\rT(k)\F_{N,N''}^{1,1} \J_{N''}^{}(kd) \r_{N''}^{}(\hx) \\
    \a_N^\rT(k)\F_{N,N''}^{1,-1} \J_{N''}^{}(kd) \r_{N''}^{}(\hx) \\
    \a_N^\rT(k)\F_{N,N''}^{1,0} \J_{N''}^{}(kd) \r_{N''}^{}(\hx)    
    \end{array} 
    \right]   \right) \nonumber\\
    &= - \sqrt{\frac{\pi}{3}}\frac{1}{c\rho_0}  \left[\begin{array}{c}
    \r_N^\rT(\hx) \J_N^{}(kd)  \A_N^\rT\F_{N,N''}^{1,1} \J_{N''}^{}(kd) \r_{N''}^{}(\hx)\\
    \r_N^\rT(\hx) \J_N^{}(kd)  \A_N^\rT\F_{N,N''}^{1,-1} \J_{N''}^{}(kd) \r_{N''}^{}(\hx)\\
    \r_N^\rT(\hx) \J_N^{}(kd)  \A_N^\rT\F_{N,N''}^{1,0} \J_{N''}^{}(kd) \r_{N''}^{}(\hx)    
    \end{array} 
    \right]  ,
\end{align}
where $\A_N^{} = \a_N^{} \a_N^\rH$.

A related operation is what is termed the \emph{energy vector} $\e$ in Ambisonic literature \cite{gerzon1992general}, which is often used as a proxy on localization performance of ambisonic decoders or multichannel panning \cite{zotter2012all, grandjean2018size}. It is related to the acoustic intensity vector in the case of incoherent summation of the plane wave signals from all directions, normalized with the total sound field energy. Its computation in terms of the coefficients of the sound field amplitude distribution at the origin is then
\begin{align}
\e(k) &= \frac{ \sint{ |a(k,\u)|^2 \u }}{ \sint{ |a(k,\u)|^2 }} = \frac{ \sint{ |a(k,\u)|^2 \u }}{ ||\a_N(k)||^2} \nonumber\\
&= \frac{1}{ ||\a_N(k)||^2} \sqrt{\frac{4\pi}{3}}   \left[\begin{array}{c}
    \sint{ (\r_N^\rT(\u)\a_N^{}(k) )^\rH(\r_N^\rT(\u)\a_N^{}(k)) R_{1,1}(\u) }\\
    \sint{ (\r_N^\rT(\u)\a_N^{}(k) )^\rH(\r_N^\rT(\u)\a_N^{}(k)) R_{1,-1}(\u)  }\\
    \sint{ (\r_N^\rT(\u)\a_N^{}(k) )^\rH(\r_N^\rT(\u)\a_N^{}(k)) R_{1,0}(\u) }     
    \end{array} 
    \right] \nonumber\\
&= \frac{1}{ ||\a_N(k)||^2} \sqrt{\frac{4\pi}{3}}   \left[\begin{array}{c}
    \sint{ (\a_N^\rH(k)\r_N^{}(\u) )(\r_N^\rT(\u)\a_N^{}(k) ) R_{1,1}(\u) }\\
    \sint{ (\a_N^\rH(k)\r_N^{}(\u) )(\r_N^\rT(\u)\a_N^{}(k) ) R_{1,-1}(\u)  }\\
    \sint{ (\a_N^\rH(k)\r_N^{}(\u) )(\r_N^\rT(\u)\a_N^{}(k) ) R_{1,0}(\u) }     
    \end{array} 
    \right] \nonumber\\    
&= \frac{1}{ ||\a_N(k)||^2} \sqrt{\frac{4\pi}{3}}   \left[\begin{array}{c}
    \a_N^\rH(k) \left(\sint{ \r_N^{}(\u)\r_N^\rT(\u) R_{1,1}(\u)  } \right) \a_N^{}(k) \\
    \a_N^\rH(k) \left(\sint{ \r_N^{}(\u)\r_N^\rT(\u) R_{1,-1}(\u)  } \right) \a_N^{}(k) \\
    \a_N^\rH(k) \left(\sint{ \r_N^{}(\u)\r_N^\rT(\u) R_{1,0}(\u) } \right) \a_N^{}(k)    
    \end{array} 
    \right]  \nonumber\\    
&= \frac{1}{ ||\a_N(k)||^2} \sqrt{\frac{4\pi}{3}}   \left[\begin{array}{c}
    \a_N^\rH(k) \F_{N,N}^{1,1} \a_N^{}(k)\\
     \a_N^\rH(k)\F_{N,N}^{1,-1} \a_N^{}(k)\\
    \a_N^\rH(k) \F_{N,N}^{1,0}  \a_N^{}(k)   
    \end{array} 
    \right].      
\end{align}

\subsection{Spherical windowing and spherical masking with applications to Ambisonics and beamforming}

In certain acoustical processing tasks a directional weighting $w(\u)$ can be applied to the directional acoustic pressure $p(\u, d)$ at some distance $d$ from the origin, or to the estimated sound field amplitude density $a(\u)$, in order to shape it in a way useful to the task. The weighting function can be derived by the application requirements and it can be fixed, or it can be signal-dependent and time-variant.

Such an operation is, for example, described in \cite{kronlachner2014spatial} in the context of processing of Ambisonic signals, termed \emph{directional loudness modification}, where a user-defined band-limited $w(\u)$ is applied to the signals to achieve directional amplification or suppression (e.g. frontal directions remain unchanged while rear ones are suppressed by 12dB, with a smooth transition in between). This is a simple application of Eq.~\ref{eq:cGmtx}, where $\a_N$ are the input Ambisonic signals and $\w_{N''}$ are the SHT coefficients of the directional weighting function. Focusing on a single channel $a_{n',m'}$ of the modified ambisonic signal vector $\a_{N'}$ we get
\begin{equation}
    a_{n',m'}(k) = \sint{a(k,\u) w(\u) R_{n',m'}(\u)} = \a_N^\rT(k) \G_{N,N''}^{n',m'} \w_{N''}^{},
    \label{eq:sphwin}
\end{equation}
with the order of the modified Ambisonic signals being $N' = N+N''$. Note that, since obtaining higher order Ambisonic signals than the input ones may be undesirable, the authors of \cite{kronlachner2014spatial} limit the output order of the transformed signals to that of the input, at the expense of a non exact modification of the product of Eq.~\ref{eq:sphwin}.

Similar operations can be described as \emph{spherical windowing}, where the window $w(\u, \u_0)$ with its main lobe centered at $\u_0$ is applied to the amplitude density $a(\u)$ in order to end up with a set of modified coefficients $a_{N'}$ that express a localized soundfield incident from around $\u_0$. Spatial analysis or reproduction of such a directionally windowed sound field focuses then mainly on the sounds incident from the spherical sector covered by the main lobe of the window. 
Eq.\ref{eq:sphwin} can be also simplified to 
\begin{align}
    \a_{N'}^\rT(k) &= \a_N^\rT(k) \left[\G_{N,N''}^{0,0} \w_{N''}(\u_0),..., \G_{N,N''}^{n',m'} \w_{N''}(\u_0),...,\G_{N,N''}^{N',N'} \w_{N''}(\u_0) \right] \nonumber\\ 
    &= \a_N^\rT(k) \W_{N,N'}^{}(\u_0)
\end{align}
where the matrix $\W_{N,N'}$ can be precomputed for a fixed $w(\u,\u_0)$. Such spherical windowing is applied in \cite{politis2016acoustic} in the context of spatial audio coding and reproduction for estimating acoustic intensity in a directionally limited sector of the sound field. Similarly, spherical windowing in \cite{sun2011optimal} is posed as an Ambisonic encoding problem from a spherical microphone array, resulting in the microphone filters required to derive the Ambisonic signals of the windowed sound field.

Another interesting and related application is described in \cite{shabtai2013generalized} aiming at simultaneous beamforming and binaural decoding of the beamformed sound field. Beamforming in this case can be described by a desired beam pattern $w(\u,\u_0)$ centered at $\u_0$ with respective SHT coefficients $\w_N(\u_0)$ matched to the sound field SH order $N$. Binaural decoding is additionally described by the spherical functions of Head-related Transfer Functions (HRTFs) $\h_{l,r}(k,\u) = [h_l(k,\u),\, h_r(k,\u)]^\rT$, where $l,r$ indicate left and right ear accordingly. The SHT coefficients of the HRTFs up to a suitable order $N'$ capturing adequately the binaural directional cues can be expressed in a matrix $\H_{N'}(k) = [\mathcal{SHT}_{N'}(h_l(k,\u)),\, \mathcal{SHT}_{N'}(h_r(k,\u))]$. The operation of obtaining the beamformed binaural signals $\b(k) = [b_l(k),\, b_r(k)]^\rT$ is then defined as
\begin{align}
    \b(k) &= \sint{ a(k,\u) w(\u,\u_0) \h_{l,r}(k,\u) } \nonumber\\
        &= \sint{ (\a_N^\rT(k)\r_N(\u)) (\w_N^\rT(\u_0)\r_N(\u)) \left(\H_{N'}^\rT(k) \r_{N'}(\u)\right) } \nonumber\\
        &= \sint{ \left(\H_{N'}^\rT(k) \r_{N'}(\u) (\r_N^\rT(\u)\w_N(\u_0))\right)\left(\sum_{n=0}^{N}\sum_{m=-n}^n a_{n,m}(k) R_{n,m}(\u)\right)  } \nonumber\\
&= \sum_{n=0}^{N}\sum_{m=-n}^n a_{n,m}(k) \H_{N'}^\rT(k) \left(\sint{ \r_{N'}(\u) \r_N^\rT(\u)R_{n,m}(\u)  }\right)\w_N(\u_0) \nonumber\\   
&= \sum_{n=0}^{N}\sum_{m=-n}^n a_{n,m}(k) \H_{N'}^\rT(k) \F_{N',N}^{n,m}\w_N(\u_0) \nonumber\\
&= \sum_{n=0}^{N}\sum_{m=-n}^n a_{n,m}(k) \tilde{\H}_{N',N}^{n,m}(k)\w_N(\u_0)
\label{eq:binbeamf}
\end{align}
where the matrices $\tilde{\H}_{N',N}^{n,m} = \H_{N'}^\rT(k) \F_{N',N}^{n,m}$ can be precomputed, while the beamforming weight vector $\w_N(\u_0)$ can remain independent, for potentially time-variant and externally determined or interactive control of the focusing direction $\u_0$. Eq.~\ref{eq:binbeamf} can be also written in a matrix form as
\begin{align}
    \b(k) =  \W_{N',N}(k, \u_0) \a_N(k)
\end{align}
where $\W_{N',N} = [\tilde{\H}_{N',N}^{0,0}(k)\w_N(\u_0), ...,\tilde{\H}_{N',N}^{n,m}(k)\w_N(\u_0),..., \tilde{\H}_{N',N}^{N,N}(k)\w_N(\u_0)]$.

\subsection{Directional array response modeling}
\label{sec:aray_model}

The far field directional response of an array of $Q$ microphones $\h(k,\u) = [h_1(k,\u),...,h_Q(k,\u)]^\rT$ to a unit-amplitude plane wave incident from $\u$, also termed \emph{array transfer function} (ATF) in literature, with respect to some origin e.g. at the center of the array or at the position of some reference sensor, are typically considered known for many spatial filtering, source localization, or signal enhancement applications. For certain idealized arrays, such as arrays of omnidirectional microphones, omnidirectional microphones mounted on an acoustically hard rigid cylindrical \cite{teutsch2007modal} or spherical baffle \cite{meyer2002highly}, or cardioid microphones oriented radially in free-field or around a spherical scatterer \cite{plessas2010microphone}, there exist analytical expressions of the ATFs. For complex arrays that may include spaced directional sensors or microphones mounted on scatterers with complex geometries, another option is direct measurement of such ATFs at a discrete number of points around the array, as is common for example in the case of HRTFs \cite{romigh2015efficient}, or through numerical simulation of the scattered field \cite{zotkin2017incident}. A useful representation of the ATFs in all cases is their SH domain one, where 
\begin{equation}
    \h(k,\u) = \H_N(k) \y_N^{}(\u)
    \label{eq:shatf}
\end{equation}
where $\H_{N}(k) = [\mathcal{SHT}_{N}(h_1(k,\u)),...,\mathcal{SHT}_{N}(h_Q(k,\u))]^\rT$. Eq.~\ref{eq:shatf} is useful for interpolation of ATFs or HRTFs measured at a discrete set of points, or for assessment of diffuse-field coherence properties of the array \cite{politis2016diffuse}. They also provide an elegant way to derive measurement-based Ambisonic encoding filters $\E_\mathrm{ls}$ from an arbitrary array as shown in \cite{jin2013design, politis2017comparing}
\begin{align}
    \E_\mathrm{ls}(k) &= \argmin_\E \sint{ ||\E(k) \h(k,\u) - \r_L(\u)||^2 }\nonumber\\
    &= \argmin_\E \sint{||\E(k) \H_N(k) \r_N(\u) - \r_L(\u)||^2 }\nonumber\\
    &= \argmin_\E \sint{||\E(k) \H_N(k) \r_N(\u) - \I_{L,N}\r_N(\u)||^2}
\end{align}
where $\mathbf{I}_{L,N} = [\mathbf{I}_L : \mathbf{0}_{(L+1)^2\times (N+1)^2-(L+1)^2}]$ is a zero-padded identity matrix that extracts the $L$th-order coefficients from a $N$th-order vector or matrix, with $L\leq N$. In practice the solution includes a Tikhonov regularization term controlled by regularization value $\lambda$ in order to avoid excessive amplification in the resulting filters, giving the solution \cite{jin2013design}
\begin{align}
    \E_\mathrm{ls}(k) &= \argmin_\E ||\E(k) \H_N(k) - \I_{L,N}||_F^2 + \lambda^2 ||\E(k) ||_F^2 \nonumber\\
    &= \I_{L,N} ^{}\H_N^\rH(k)\left( \H_N^{}(k)\H_N^\rH(k) + \lambda^2\I_M^{} \right) \nonumber\\
     &= \H_L^\rH(k)\left( \H_N^{}(k)\H_N^\rH(k) + \lambda^2\I_M^{} \right).
\end{align}
Such a formulation of the SH encoding typically results in filters with less temporal spread and pre-ringing than similar least squares formulations directly using the discrete measurements of ATFs \cite{politis2017comparing}. 

Regarding modeling of the array coefficient matrix $\H_N(k)$, apart from the aforementioned analytical cases, a few more general cases of practical interest can be defined in a straightforward manner. Let us assume an array of $Q$ spaced microphones, each with its own complex known directivity function $d_q(k,\u)$ measured in the coordinate system of each individual microphone, as is typically the case with directivity data coming from the manufacturer of a certain microphone. Now let us assume that each of the microphones is centered at a position $\x_q$ and at an orientation described by a rotation matrix $\R_\mathrm{xyz}^{(q)}$. The array ATF component $h_q(k,\u)$ for that microphone is then, including the local microphone rotation and the phase delay term due to the spacing of the microphone from the array origin
\begin{align}
    h_q(k,\u) = d_q(k, (\R_\mathrm{xyz}^{(q)})^{-1} \u) e^{ik\u^\rT \x_q}.
\end{align}
The SH coefficients of the $q$th microphone with respect to the array coordinate system are then
\begin{align}
    h_{n,m}^{(q)}(k) &= \sint{ d_q(k, (\R_\mathrm{xyz}^{(q)})^{-1} \u) e^{ik\u^\rT \x_q}R_{n,m}(\u)} \nonumber\\
     &= \sint{ \left(\d_{N'}^\rT(k) \R_\mathrm{shd}^{(q)} \r_{N'}^{}(\u) \right) \left( \r_{N''}^\rT(\u) \J_{N''}^{}(kd) \r_{N''}^{}(\hx_q) \right) R_{n,m}(\u)} \nonumber\\
&= \d_{N'}^\rT(k) \R_\mathrm{shd}^{(q)} \left(\sint{\r_{N'}^{}(\u)\r_{N''}^\rT(\u) R_{n,m}(\u)} \right)\J_{N''}^{}(kd) \r_{N''}^{}(\hx_q)   \nonumber\\  
&= \d_{N'}^\rT(k) \R_\mathrm{shd}^{(q)} \F_{N',N''}^{n,m} \J_{N''}^{}(kd) \r_{N''}^{}(\hx_q)  
\end{align}
with $\R_\mathrm{shd}^{(q)}$ being the SH domain rotation matrix of the standard rotation matrix $\R_\mathrm{xyz}^{(q)}$ \cite{blanco1997evaluation}, and $\d_{N'}(k) = r\mathcal{SHT}_{N'}[d_q(k,\u)]$ the SH coefficients of the original non-translated and unrotated microphone directivity function.

\subsection{Sound field and inter-sensor correlation modeling under isotropic and anisotropic diffuse conditions}

In various acoustical signal processing applications, a statistical model of the sound field is useful, with the most widespread example being an isotropic directionally incoherent distribution of plane-waves of equal signal variance (or power) across directions. We can model such a sound field through a directional amplitude density $d(k,\u)$ with second order statistics
\begin{align}
    \mathbb{E}[d(k, \u)d^*(k, \u')] = \begin{cases}
  P_d(k)  & \text{if  } \u = \u' \\
  0 & \text{if  } \u \neq \u'.
\end{cases}
\end{align}
where $\mathbb{E}[\cdot]$ denotes statistical expectation. Hence, an isotropic diffuse sound field is characterized by a constant directional power spectral density. A well known result is that the sound field coefficients $\d_N(k) = r\mathcal{SHT}_N[d(k,\u)]$ of a diffuse isotropic field are also incoherent between different orders and degrees, as exhibited in their spatial covariance matrix (SCM) being identity \cite{politis2016diffuse}
\begin{align}
    \mathbb{E}[\d_N^{}(k)\d_N^\rH(k)] &= \mathbb{E}\left[\left( \sint{d(k,\u)\r_N^{}(\u)} \right) \left( \sintt{ d(k,\u')\r_N^{}(\u') } \right)^\rH \right]\nonumber\\
    &= \mathbb{E}\left[ \sint{\sintt{d(k,\u)\r_N^{}(\u) d^*(k,\u')\r_N^\rT(\u') }}  \right]\nonumber\\
    &= \sint{\sintt{\mathbb{E}\left[d(k,\u)d^*(k,\u')\right]\r_N^{}(\u) \r_N^\rT(\u') }}  \nonumber\\
&= \sint{P_d(k) \r_N^{}(\u) \r_N^\rT(\u) }  = P_d(k)\sint{ \r_N^{}(\u) \r_N^\rT(\u) }  \nonumber\\
&= P_d(k)\I_N \label{eq:isotropic_ambi}
\end{align}
where $\I_N$ is the $(N+1)^2\times (N+1)^2$ identity matrix and the SH orthonormality property $\sint{ \r_N^{}(\u) \r_N^\rT(\u) } =  \I_N$ is used in the last step.
Having an array around the origin with ATFs given by $\h(k,\u)$, and the respective ATF coefficient matrix $\H_N(k)$, the array signals under isotropic diffuse conditions are $\z(k) = \H_N(k)\d_N(k)$ with their SCM being
\begin{align}
    \mathbb{E}[\z(k)\z^\rH(k)] &=\H_N(k)\mathbb{E}\left[\d_N^{}\d_N^\rH\right]\H_N^\rH(k)\nonumber\\
    &= P_d(k)\H_N^{}(k)\H_N^\rH(k). \label{eq:isotropic_array}
\end{align}
The diffuse-field SCM of microphone arrays (or its power-normalized diffuse coherence matrix version) is an important component of adaptive and non adaptive spatial filtering \cite{bitzer2001superdirective, mccowan2003microphone}. In the simplest case of spaced omnidirectional microphones it reduces to the well known result of a sinc function of the wavenumber-distance product of pairs of microphones \cite{mccowan2003microphone}. One can arrive on the same result by modeling the SH-ATF matrix using the plane wave expansion of Eq.\ref{eq:pwe}. For more complex arrays of directional microphones the SH-ATF matrix can be constructed as described in Sec.\ref{sec:aray_model}
and detailed in \cite{politis2016diffuse}.

Another case of interest is a sound field with diffuse properties that may however not be isotropic, meaning that its directional power spectral density is not constant with direction but varies smoothly, i.e.
\begin{align}
    \mathbb{E}[d(k, \u)d^*(k, \u')] = \begin{cases}
  P_d(k,\u)  & \text{if  } \u = \u' \\
  0 & \text{if  } \u \neq \u'.
\end{cases} \label{eq:anisotropic_dpd}
\end{align}
In array processing literature such as field model is sometimes termed as a colored noise field \cite{friedlander1995direction}, but it is also useful for modeling anisotropic reverberant fields \cite{bastine2022time}, incoherently spread and extended sources \cite{mccormack2021rendering}, diffuse reflections and scattering distributions \cite{valaee1995parametric}. The SCM of the sound field coefficients under such an anisotropic directional power distribution, similar to the derivation of Eq.~\ref{eq:isotropic_ambi}, becomes
\begin{align}
    \mathbb{E}[\d_N^{}(k)\d_N^\rH(k)] 
&= \sint{P_d(k,\u) \r_N(\u) \r_N^\rT(\u) }. \label{eq:anisotropic1_ambi}
\end{align}
Assuming that the directional power spectral density $P_d(k,\u)$ is smoothly varying and directionally band limited to some order $N'$, it can be expressed itself by a vector of SH coefficients $\p_{N'}(k) = r\mathcal{SHT}_{N'}[P_d(k,\u)]$. Expanding Eq.~\ref{eq:anisotropic1_ambi} accordingly
\begin{align}
    \mathbb{E}[\d_N^{}(k)\d_N^\rH(k)] 
&= \sint{P_d(k,\u) \r_N^{}(\u) \r_N^\rT(\u) } \nonumber\\
&= \sint{\left( \sum_{n'=0}^{N'}\sum_{m'=-n'}^{n'} p_{n',m'}(k) R_{n',m'}(\u)\right)\r_N^{}(\u) \r_N^\rT(\u) } \nonumber\\
&= \sum_{n'=0}^{N'}\sum_{m'=-n'}^{n'} p_{n',m'}(k) \sint{R_{n',m'}(\u)\r_N^{}(\u) \r_N^\rT(\u) } \nonumber\\
&= \sum_{n'=0}^{N'}\sum_{m'=-n'}^{n'} p_{n',m'}(k) \F_{N,N}^{n',m'}.
\end{align}
Similarly to Eq.~\ref{eq:isotropic_array}, an array with ATFs given by $\h(k,\u)$ and the respective ATF coefficient matrix $\H_N(k)$, inside the anisotropic diffuse field characterized by $\p_{N'}(k)$ captures signals with SCM
\begin{align}
    \mathbb{E}[\z(k)\z^\rH(k)] &= \H_N^{}(k)\left(\sum_{n'=0}^{N'}\sum_{m'=-n'}^{n'} p_{n',m'}(k) \F_{N,N}^{n',m'}\right)\H_N^\rH(k) \nonumber\\
    &= \sum_{n'=0}^{N'}\sum_{m'=-n'}^{n'} p_{n',m'}(k) \H_N^{}(k)\F_{N,N}^{n',m'}\H_N^\rH(k).
\end{align}
Note that for a certain array the matrices $\H_N^{}(k)\F_{N,N}^{n',m'}\H_N^\rH(k)$ are independent of the directional power distribution and can be precomputed up to a maximum order $N'$ for any such anisotropic field of that modeling order or less.

Finally, a complementary view on modeling the statistics of a sound field under isotropic or anisotropic diffuse conditions is that of covariances between the sound field coefficients $\d_N$ at different points. Let as assume isotropic diffuse conditions, with the first set of coefficients captured at the origin, while the second one at position $\x$. The SCM between them is then
\begin{align}
    \mathbb{E}[\d_N^{}(k,\mathbf{0})\d_N^\rH(k,\x)] &= \mathbb{E}\left[\left( \sint{d(k,\u)\r_N(\u)} \right) \left( \sintt{ d(k,\u')e^{ik{\u'}^\rT\x}\r_N(\u') } \right)^\rH \right]\nonumber\\
    &= \mathbb{E}\left[ \sint{\sintt{d(k,\u)\r_N^{}(\u) d^*(k,\u')e^{-ik{\u'}^\rT\x}\r_N^\rT(\u') }}  \right]\nonumber\\
    &= \sint{\sintt{\mathbb{E}\left[d(k,\u)d^*(k,\u')\right]\r_N(\u) e^{-ik{\u'}^\rT\x} \r_N^\rT(\u') }}  \nonumber\\
&= P_d(k) \sint{ e^{ik\u^\rT(-\x)} \r_N(\u) \r_N^\rT(\u) }  \nonumber\\ 
&= P_d(k) \sint{\left( \sum_{n=0}^{N'} \sum_{m'=-n'}^{n'} 4\pi i^{n'} j_{n'}(kd)  R_{n',m'}(\u) R_{n',m'}(-\hx) \right) \r_N(\u) \r_N^\rT(\u) }  \nonumber\\ 
&= P_d(k) \sum_{n=0}^{N'} \sum_{m'=-n'}^{n'} 4\pi i^{n'} j_{n'}(kd) R_{n',m'}(-\hx) \sint{ \r_N(\u) \r_N^\rT(\u) R_{n',m'}(\u) }  \nonumber\\
&=4\pi P_d(k) \sum_{n=0}^{N'} \sum_{m'=-n'}^{n'}  i^{n'} j_{n'}(kd) R_{n',m'}(-\hx) \F_{N,N}^{n',m'}. \label{eq:isotropic_spacedambi}
\end{align}
The anisotropic directional distribution case of Eq.~\ref{eq:anisotropic_dpd} can be similarly derived, but results in more complicated expressions since it involves products of four spherical harmonics. Such products can be computed through products of Gaunt coefficients. Following the derivation of Eq.~\ref{eq:isotropic_spacedambi} and using Eq.~\ref{eq:anisotropic_dpd} and Eq.~\ref{eq:2shprod}
\begin{align}
    \mathbb{E}[\d_N^{}(k,\mathbf{0})\d_N^\rH(k,\x)] 
&=  \sint{P_d(k,\u) e^{ik\u^\rT(-\x)} \r_N(\u) \r_N^\rT(\u) }  \nonumber\\ 
&= \sint{\left( \sum_{n''=0}^{N''}\sum_{m''=-n''}^{n''} p_{n'',m''}(k) R_{n'',m''}(\u)\right) \cdot \nonumber\\
&\cdot \left( \sum_{n=0}^{N'} \sum_{m'=-n'}^{n'} 4\pi i^{n'} j_{n'}(kd)  R_{n',m'}(\u) R_{n',m'}(-\hx) \right) \r_N(\u) \r_N^\rT(\u) }  \nonumber\\ 
&= 4\pi \sint{\left( \sum_{n''=0}^{N''}\sum_{m''=-n''}^{n''} \sum_{n=0}^{N'} \sum_{m'=-n'}^{n'} p_{n'',m''}(k) i^{n'} j_{n'}(kd) R_{n',m'}(-\hx) R_{n'',m''}(\u) R_{n',m'}(\u)\right) \cdot \nonumber\\
&\cdot \r_N(\u) \r_N^\rT(\u) }  \nonumber\\ 
&= 4\pi \sint{\left( \sum_{n''=0}^{N''}\sum_{m''=-n''}^{n''} \sum_{n=0}^{N'} \sum_{m'=-n'}^{n'} p_{n'',m''}(k) i^{n'} j_{n'}(kd) R_{n',m'}(-\hx)\right. \cdot \nonumber\\
&\cdot \left. \sum_{n'''=|n'-n''|}^{n'+n''} \sum_{m'''=-n'''}^{n'''} F_{n',m',n'',m''}^{n''',m'''}R_{n''',m'''}(\u)\right)\r_N(\u) \r_N^\rT(\u) }  \nonumber\\ 
&= 4\pi \sum_{n''=0}^{N''}\sum_{m''=-n''}^{n''} \sum_{n=0}^{N'} \sum_{m'=-n'}^{n'} p_{n'',m''}(k) i^{n'} j_{n'}(kd) R_{n',m'}(-\hx) \cdot \nonumber\\
&\cdot \sum_{n'''=|n'-n''|}^{n'+n''} \sum_{m'''=-n'''}^{n'''} F_{n',m',n'',m''}^{n''',m'''} \sint{ R_{n''',m'''}(\u)\r_N(\u) \r_N^\rT(\u) }  \nonumber\\ 
&= 4\pi \sum_{n''=0}^{N''}\sum_{m''=-n''}^{n''} \sum_{n=0}^{N'} \sum_{m'=-n'}^{n'} p_{n'',m''}(k) i^{n'} j_{n'}(kd) R_{n',m'}(-\hx) \cdot \nonumber\\
&\cdot \sum_{n'''=|n'-n''|}^{n'+n''} \sum_{m'''=-n'''}^{n'''} F_{n',m',n'',m''}^{n''',m'''} \F_{N,N}^{n''',m'''}
\end{align}
Note that the last summation can be simplified further depending on relations between degrees for non-zero coefficients \cite{homeier1996some}.

\section{Software implementation}

Example routines for the computation of Gaunt coefficients in the matrix form $\G_{N',N''}^{n,m}$ of Eq.~\ref{eq:cGmtx} for complex SHs and $\F_{N',N''}^{n,m}$ of Eq.~\ref{eq:rGmtx} for real SHs are provided by the author using Matlab programming language at \url{https://github.com/polarch/Spherical-Harmonic-Transform}. The entries of the matrix $\G_{N',N''}^{n,m}$ for complex SHs are computed through Wigner-3j symbols using Eq.~\ref{eq:Cruzan}. Wigner-3j symbols are computed directly through Racah's formula \cite[\href{https://dlmf.nist.gov/34.2}{(34.2.4)}]{NIST:DLMF}. To speed up computation of large factorials, a faster alternative routine is provided that uses Stirling's approximation\footnote{\url{https://mathworld.wolfram.com/StirlingsApproximation.html}}. 

Gaunt coefficient matrices $\F_{N',N''}^{n,m}$ for real SHs are computed by transforming the matrices $\G_{N',N''}^{n,m}$ for complex SHs using Eq.~\ref{eq:cGmtx2rGmtx}. A pre-computed lookup table of both $\G_{N',N''}^{n,m}$ and $\F_{N',N''}^{n,m}$ is included in the code repository for $N'=N''=30$ and $n=1,...,30$ indexed in the third dimension. These tables allow computation of products of two spherical functions with combined degree up to $N'+N''=30$.

\bibliographystyle{ieeetr} 
\bibliography{refs} 

\end{document}